\documentclass[prl,twocolumn,groupedaddress,showpacs]{revtex4}
\usepackage[]{umlaut}
\usepackage[]{amssymb}
\usepackage[]{graphicx}

\begin{document}

\bibliographystyle{apsrev}

\title{Electron-phonon coupling and its evidence in the photoemission spectra of lead}
\author{F. Reinert}
\email[corresponding author. Email: ]{f.reinert@physik.uni-saarland.de}
\author{B. Eltner}
\author{G. Nicolay}
\author{D. Ehm}
\author{S. Schmidt}
\author{S. Hüfner}
\affiliation{Universität des Saarlandes, Fachrichtung 7.2 --- Experimentalphysik, D-66041 Saarbrücken, Germany}

\date{\today}

\begin{abstract}
We present a detailed study on the influence of strong electron-phonon
coupling to the photo\-emission spectra of lead. Representing the 
strong-coupling regime of superconductivity, the spectra of lead show
characteristic features that demonstrate the correspondence of
physical properties in the normal and the superconducting state, as predicted by the Eliashberg theory. These features
appear on an energy scale of a few meV and are accessible for
photoemission only by using modern spectrometers with high resolution
in energy and angle.  
\end{abstract}

\maketitle

A striking evidence of electron-phonon interaction in solids is
the existence of superconductivity. Bardeen, Cooper, and Schrieffer
showed \cite{bcs57} that even a weak electron-phonon coupling is
able to condense two electrons to a so-called
Cooper-pair, which is the basic
prerequisite of the superconducting ground
state and the key feature of the BCS model. The BCS theory describes successfully the superconducting
properties of many solids, e.g. Al, V, or V$_3$Si, where the
electron-phonon coupling is sufficiently weak. Other conventional 
superconductors, e.g. Pb, Hg, or Nb$_3$Ge, show significant quantitative
and qualitative deviations from the predictions of the BCS model
\cite{carbotte90}. 
These systems are usually classified as
{\em strong-coupling superconductors}.

The theoretical approach for the explanation of strong-coupling
superconductors is based on the Eliashberg equations, with the
coupling function $\alpha^2F$ as the central property
\cite{eliashberg60}. This so-called Eliashberg function can be
calculated e.g. by first principle methods from the electronic wave
functions, the phonon density of states, and the electron-phonon
coupling between two Bloch states. However, $\alpha^2F$ 
is a much more universal quantity and determines
also the influence of the electron-phonon interaction on the normal
state properties, e.g. the electrical
resistivity, the electronic heat capacity,
and --- as a spectroscopic feature --- the electron-phonon
contribution $\Gamma_{el-ph}$ to the intrinsic quasi-particle
linewidth, which can be determined
e.g. by photoemission spectroscopy. In the Green's function method, the influence of
electron-phonon interaction is expressed as a contribution
to the complex self energy of the conduction electrons $\Sigma_{el-ph}$, which
can be calculated from the Eliashberg function $\alpha^2F$ \cite{grimvall69}. The real part
describes the induced band
renormalization, where the
imaginary part gives the quasi-particle line width,
equivalent to the reciprocal hole lifetime $\tau$.
In particular, at very small energies (i.e. close to the Fermi level), the real part
of $\Sigma_{el-ph}$ is linear in energy, and 
$\lambda=-\delta\Re{\rm e}\Sigma_{el-ph}(\omega)/\delta\omega$ at
$\omega=0$ is usually called the mass enhancement factor. In addition,
$\lambda$ defines the slope of the temperature dependence of the quasi-particle
linewidth $\Gamma_{el-ph}=2\Im{\rm m}\Sigma_{el-ph} \approx 
2\pi\lambda k_BT$ for temperatures
above the Debye temperature $\Theta_D$.




Photoemission spectroscopy (PES) is a versatile
experimental method to study the electronic structure of
solids and has been applied to many high-$T_c$ materials \cite{damascelli03,campuzano02}
and conventional superconductors \cite{grioni91,chainani00,bcs_reinert00,tsuda01,yokoya01}.  
Electron-phonon coupling in metallic systems
has been studied in detail by PES on
low-dimensional electronic states at surfaces, e.g.
Shockley-states
\cite{paniago95,mcdougall95,balasubramanian98,hengsberger99b,valla99,lashell00,eiguren02}
or quantum-well states \cite{paggel99,takahashi99}. PES
investigations on electron-phonon effects in
three-dimensional solids, however, are rare. One
particularly interesting three-dimensional model system is Pb, which
among the conventional superconductors has quite un-conventional
physical properties. 


\begin{figure}[t]
~
  \begin{center}
    \includegraphics[width=7cm]{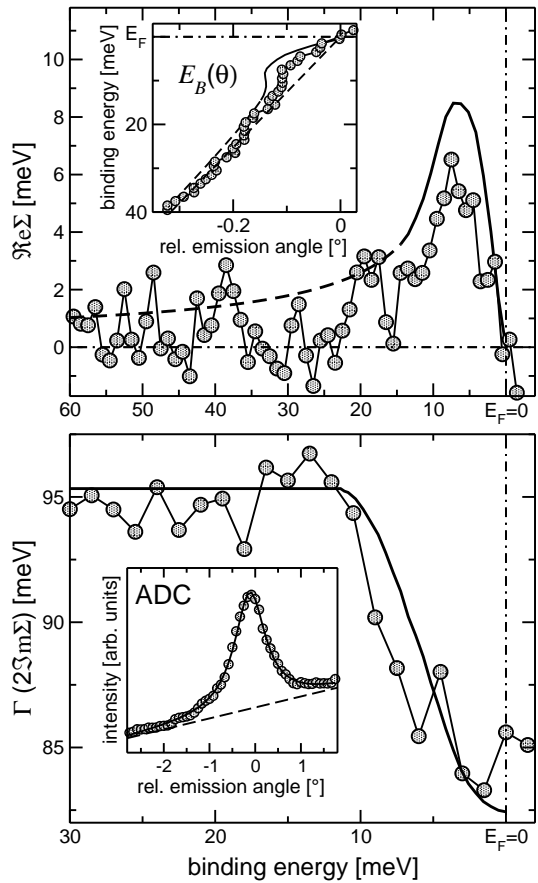}
    \caption{Real and imaginary part of the electron-phonon self
energy $\Sigma$, experimental data (circles) and calculated from Ref.
\cite{grimvall69}(full lines). Upper inset: measured band
dispersion close to $E_F$ in
        the normal phase ($T\!\approx\!8$~K); the
        dashed line gives the result of a linear fit to the data points at
        higher energies ($E_B\!>\!20$~meV). The experimental data points have
        been determined from the maximum position of a Lorentzian fit to the
        angular distribution curves (ADC, see lower inset with ADC at
        $E_B\!=\!10.5$~meV).} 
    \label{fig:dispersion}
  \end{center}
\end{figure}

In the case of Pb the two most relevant energy scales on which the
electron-phonon features appear in the spectra, the Debye
energy $\hbar\omega_D=k_B\Theta_D$ and the
gap width $\Delta_0$ at $T=0$, define prerequisites that can be met
by PES experiments today ($\hbar\omega_D\!=\!7.6$~meV,
$\Delta_0\!=\!1.4$~meV). The instrumental resolution
must be at least of the same
order of magnitude to give usable spectral information on these
features, i.e. $\Delta E \lesssim 3$~meV. Furthermore, if
one 
wants to measure the band renormalization on single-crystalline
samples, the angular resolution must be sufficient to resolve the
band dispersion of states close to the Fermi level. Since a few years,
photoelectron spectrometers fulfil these
requirements and PES has become a unique method yielding access to electron-phonon features in
both the density of states and the $k$-resolved electronic band structure.

The photoemission data presented here have been measured using He~{\sc I}$_\alpha$ radiation
from a monochromatized helium discharge lamp ($h\nu = 21.22$~eV) with an
energy resolution of  $\lesssim3$~meV. The angular resolution was typically
$\approx 0.3^\circ$ and the sample temperature could be varied from
room temperature down to
about $4.5$~K (see Ref.\ \cite{bcs_reinert00,cuagau_reinert01} for details of
the experimental setup). 
The samples were polycrystalline and single-crystalline pellets (Surface Preparation Laboratory, Netherlands),
thermally connected to the sample holder by soldered
indium. Clean single-crystalline surfaces were prepared {\it in situ} by careful Ar sputtering
and subsequent annealing
at $\lesssim150^\circ$C.

\begin{figure}[htb]
~
  \begin{center}
    \includegraphics[width=7.8cm]{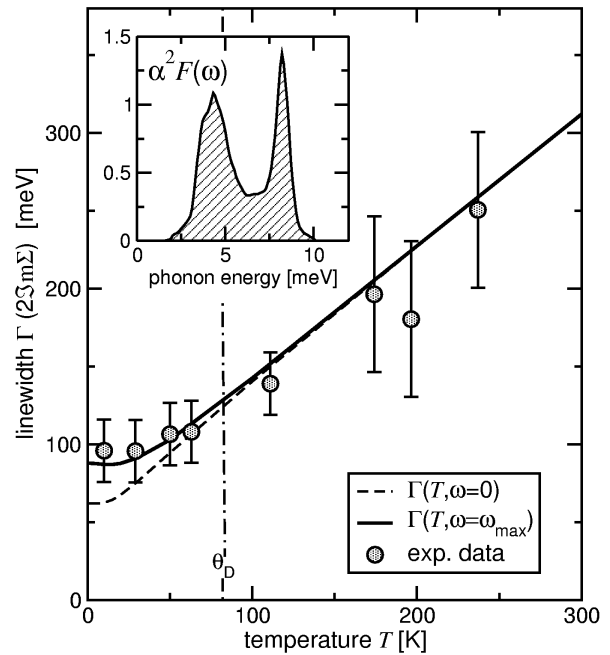}
    \caption{Temperature dependence of the lifetime width $\Gamma$ for
energies $\omega\gtrsim\omega_D$ (circles). 
The dashed and the solid line represent the calculated
electron-phonon contribution at $E_F\!=\!0$ and
$E\!=\!\hbar\omega_{\rm max}\!=\!12$~meV, shifted vertically to match
the experimental data. The inset shows $\alpha^2F(\omega)$ as used for the calculation (from Ref.\ \cite{mcmillan64}). }
    \label{fig:Gamma_tempdep}
  \end{center}
\end{figure}


On such a surface, one is able to investigate the electronic band
structure of states close to the Fermi level by angular resolved
PES (ARUPS).  The inset in the upper panel of Fig.\ \ref{fig:dispersion} gives
a volume band dispersion $E_k$ close to the Fermi level on a 
Pb(110) surface in the normal state ($T\!=8$~K).
With He~{\sc I}$_\alpha$, this band, forming the locally tubular Fermi surface of
Pb, can be reached slightly off the high-symmetry direction
$\overline{\Gamma K}$ \cite{horn84}, or, more precisely, at
$k_{\parallel,x}=0.82$\AA$^{-1}$ along
$\overline{\Gamma K}$ and
$k_{\parallel,y}=0.11$\AA$^{-1}$ parallel to $\overline{\Gamma
X}$. The data points (circles)
were determined from the position of the peak maxima in the angular
distribution curves (ADC) at the
energies close to the Fermi level $E_F$ (see lower inset).  At higher energies,
i.e. significantly higher than the Debye energy $\hbar\omega_D=7.6$~meV, the
dispersion can be approximated quite well by a linear function of the
momentum $k$, representing the dispersion $\epsilon_k$ with no
renormalization.
However, 
below $E\approx12$~meV the band shows the well known correction due
to electron-phonon interaction, which is defined
by the real part of the self energy $\Re{\rm
e}\Sigma(E_k)=E_k-\epsilon_k$. The upper panel shows the experimental
result ($T=8$~K) for the deviation from the linear dispersion
compared to the calculated $\Re{\rm
e}\Sigma$ from Ref.\ \onlinecite{grimvall69} at $T=11$~K.
The comparison shows a good qualitative agreement, in particular the position of the maximum at $\approx7$~meV and the
slope at $E_F$ (given by the coupling constant $\lambda=1.55$) are the same for experiment and theory, within the
experimental uncertainties.  The difference in maximum
height can be explained by the finite experimental resolution. 
Up to now, such a renormalized quasi-particle dispersion
has been observed only for
low-dimensional systems, e.g. the Shockley-state on Mo(110)
\cite{valla99} or high-$T_c$ superconductors \cite{damascelli03,campuzano02}.

\begin{figure}[ht]
  \begin{center}
    \includegraphics[width=6.8cm]{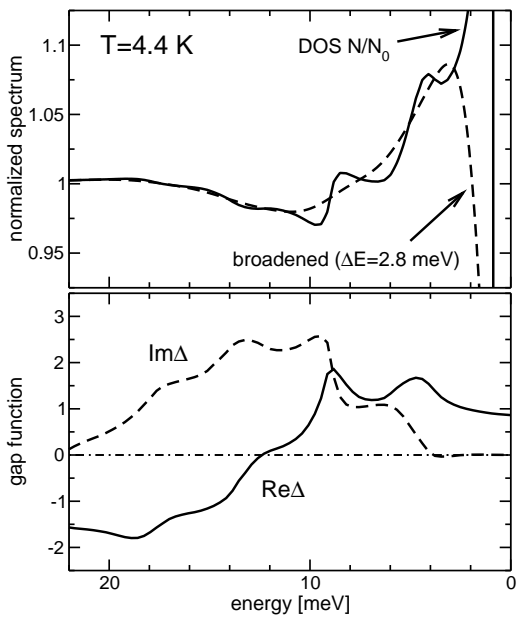}
    \caption{Lower panel: real and imaginary part of the gap function,
        interpolated at $T\!=\!4.4$~K from the calculated gap functions at $T\!=\!0$  
        and $0.98T_c$ given in Ref.\ \cite{scalapino65}. Upper panel: resulting
        spectrum, raw normalized density of states $N/N_0$ (solid line) and broadened by a
        convolution with a Gaussian (FWHM $\Delta E\!=\!2.8$~meV) to
        simulate the resolution function.} 
    \label{fig:deltas}
  \end{center}
\end{figure}

By an analysis of the ADC linewidth as a function of the binding energy
one can extract the energy dependence of the imaginary part $\Im{\rm
m}\Sigma$ of the self energy (see
e.g. Refs.~\onlinecite{hengsberger99b,valla99}). 
The lower panel of Fig.~\ref{fig:dispersion} shows the experimental
energy dependence in comparison to $\Im{\rm m}\Sigma(E_k)$ from Ref.\
\onlinecite{grimvall69}. 
Since Pb is a three-dimensional system, the contribution of the
final state to the total linewidth can not be neglected \cite{smith93}.
This additional contribution is not related to
electron-phonon interaction and therefore does not show an energy or
temperature dependence in the investigated range.
Thus, it is reasonable to simply
shift the theoretical result by 82~meV to match the photoemission data.
Apart from this offset and experimental limitations, the agreement
between experimental result and theory is good: below the maximum
phonon energy 
$\hbar\omega_{max}\approx10$~meV (see inset of Fig.\
\ref{fig:Gamma_tempdep}) there is a continous increase of the
linewidth up to a net change of $\approx13$~meV. As known from the
Debye model, above $\hbar\omega=10$~meV the electron-phonon
contribution to the linewidth remains constant.

The photoemission linewidth is also characteristically dependent on the temperature.
Fig.~\ref{fig:Gamma_tempdep} shows the experimental linewidth from
ADCs for $\omega\gtrsim\omega_D$ (circles);
the dashed and the solid line give the
theoretical results at $E_F$ and $E_B=\hbar\omega_{max}$,
respectively. 
Obviously, the calculated temperature
dependence describes the experiment quite well. For the whole
investigated temperature range, the theoretical curve lies within the
error bars of the experimental values and, in particular, for high temperatures the
experimental linewidth
follows the predicted linear behavior given by $\Gamma_{el-ph}=2\pi\lambda
k_BT$, again with $\lambda\!=\!1.55$. This
result confirmes that the contribution of the final state to the total
photoemission linewidth does not considerably depend on the temperature.


\begin{figure}[t]
  \begin{center}
    \includegraphics[width=6.8cm]{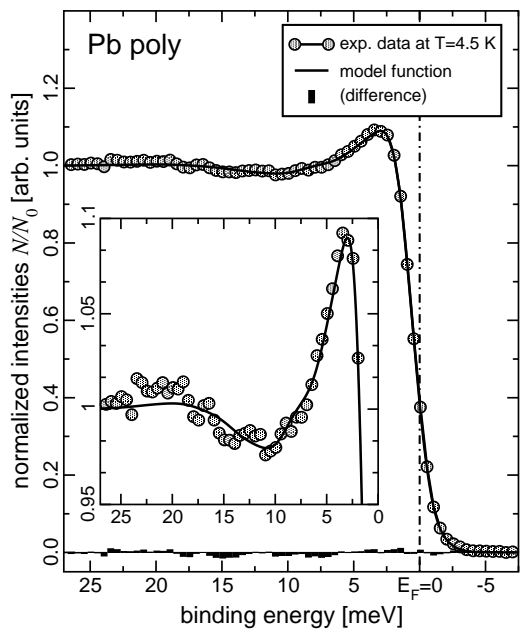}
    \caption{Comparison of the modelled function from Fig.\
\ref{fig:deltas} (solid line) with the experimental data
(circles). The black bars at the x-axis indicate the difference
between experiment and theory. The inset shows a blow-up of the
peak and dip structure.}
    \label{fig:fit}
  \end{center}
\end{figure}

Obviously, there is clear evidence for the strong electron-phonon
interaction in the photoemission spectra of Pb in the normal state.
In addition, there are characteristic spectral features in the
superconducting state of Pb: a
peak-and-dip structure, that has been
experimentally observed first by tunneling techniques 
\cite{giaever62} and reproduced recently in photoemission
data \cite{chainani00}.
Theoretically, the density of states of a strongly
coupled superconductor can be fully described by an energy and temperature
dependent, complex gap function $\Delta(E,T)$ given by the
Eliashberg theory \cite{schrieffer63,scalapino66}. The imaginary
part of $\Delta$ is the consequence of the damping of the quasi-particle
excitations caused by the electron-phonon interaction. With this gap function, the
spectral density of states of a strongly coupled superconductor is 
described by
$N(E,T)/N_0 = \Re{\rm e}\left\{ E/\sqrt{E^2 - \Delta^2(E,T)}\right\}$,
giving as a special case the dip-less BCS density of states when $\Delta$
chosen real and constant in $E$.


A theoretical gap function for Pb at finite temperatures can be found in
Ref.~\onlinecite{scalapino65}, which gives the energy dependence for $\Delta(E,T)$ explicitly
for two temperatures $T_1=0$
and $T_2=0.98T_c$. To obtain the gap function at the
temperature of the present experiment ($T=4.4$~K), we simply
interpolated between these two pairs of curves 
$\Delta(E,T)=(1-\alpha(T))\Delta(E,T_1)+\alpha(T)\Delta(E,T_2)$, considering
 the temperature dependence of the BCS gap \footnote[29]
{The weighting factor $\alpha(T)$ was determined by
the ratio of the respective BCS gaps at $T$ and $T=T_2$, resulting
in $\alpha(T)=(\Delta_0-\Re{\rm e}\{\Delta(0,T)\})/(\Delta_0-\Re{\rm
  e}\{\Delta(0,T_2)\})$.}.

We model the photoemission spectrum by convoluting the result with a
Gaussian to describe the broadening from the finite experimental
resolution. The interpolated gap function and the resulting spectrum for
$T=4.4$~K are displayed in Fig.~\ref{fig:deltas}.

Fig.~\ref{fig:fit} shows a comparison of the model function with the
experimental data at $T=4.4$~K, normalized to unity at $E=27$~meV
where the spectrum becomes flat; the inset shows a blow-up of the
range of the dip. The solid line represents the model function from 
Fig.~\ref{fig:deltas}, the bars at the x-axis represent the difference
to the normalized experimental spectrum. The agreement between
experiment and model is striking. Without using any free 
fit parameter, all important features ---the position of the gap edge, the intensity ratio between peak and
dip, and the spectral shape in the displayed energy range--- are
perfectly described by the model function.  

The most important feature of the spectrum in Fig.~\ref{fig:fit} is the
drop of the intensity around $11$~meV below unity, which is also lower
than the
BCS intensity and the normal state DOS in this energy range. Choosing a simple Einstein-like model phonon at energy
$\hbar\omega_E$, it was demonstrated \cite{scalapino66} that 
the dip in the DOS appears approximately at a binding energy slightly beyond
$\hbar\omega_E\!+\!\Delta_0$ --- corresponding to the maximum position in
$\Im{\rm m}\Delta$. This is in accordance with our observation 
where the dip appears at an energy slightly higher
than $\hbar\omega_D\!+\!\Delta_0=9$~meV.
In the BCS model, the superconducting gap width $\Delta_0$ at $T=0$ is related
to the transition temperature $T_c$ by the dimensionless parameter
${2\Delta_0}/{k_BT_c}=3.50$.
 This parameter amounts to 4.497 for Pb ($T_c\!=\!7.19$~K), much
larger than the BCS value; Al ($T_c\!=\!1.18$~K) as 
the prototype BCS material has 
3.535 \cite{carbotte90}. 
This means, the gap width of Pb is larger than for a
BCS like superconductor with the same $T_c$. As a
consequence, the
characteristic weak photoemission structure from the thermally occupied
singularity above $E_F$, which can be observed e.g. at
$T\approx0.6T_c$ for V$_3$Si
\cite{bcs_reinert00}, is not resolvable  because the Fermi-Dirac
distribution suppresses the spectral intensity at energies farther away from
the Fermi level.

In conclusion, the high-resolution photoemission data presented in this paper
clearly demonstrate the influence of the strong electron-phonon coupling
on the electronic structure of lead in both the normal and superconducting
state. The experimental superconducting density of states, the 
renormalized band dispersion, the energy and the temperature dependence of the
quasi-particle linewidth are in very good agreement with theoretical
calculations based on the Eliashberg theory and represent consistently
--- and even quantitatively ---
the electron-phonon corrections in spectroscopic data. 
We hope that these results may also help to reveal the driving
mechanisms behind the unconventional superconductors, like high-$T_c$
systems, superconducting heavy-fermion compounds and even
superconducting organic materials.

This work was supported by the Deutsche
Forschungsgemeinschaft (grant nos.\ Hu 149-19-1 and SFB 277).


\end{document}